\documentclass[final,english]{bullsrsl}[2022/06/15]

\usepackage{hyperref}

\usepackage[latin1]{inputenc}
\usepackage[T1]{fontenc}

\usepackage{natbib} 
\usepackage{graphicx,subfigure}

\newcommand{\spo}{$\omega$\,+\,$\Omega$\,} 
\newcommand{\smo}{$\omega$\,-\,$\Omega$\,} 
\newcommand{\psmo}{P$_{\omega\,-\,\Omega}$\,} 
\newcommand{\po}{P$_{\Omega}$\,} 
\newcommand{\ob}{${\Omega}$\,} 
\newcommand{\ps}{P$_{\omega}$\,} 
\newcommand{\s}{${\omega}$\,} 

\begin{document}
\title{ X-ray observations of the Intermediate Polar TX Col \footnote{Based on X-ray data obtained from Suzaku, Chandra and Swift}}

\author[affil={1}, corresponding]{Jeewan Chandra}{Pandey}
\author[affil={1}]{Nikita}{Rawat}
\author[affil={1}]{Srinivas M.}{Rao}
\author[affil={2}]{Arti}{Joshi}
\author[affil={3}]{Sadhana}{Singh}
\affiliation[1]{Aryabhatta Research Institute of Observational Sciences (ARIES), Nainital -263001, India}
\affiliation[2]{Pontificia Universidad Cat\'{o}lica de Chile, Av. Vicu\~{n}a Mackenna 4860, 782-0436 Macul, Santiago, Chile}
\affiliation[3]{Astronomy \& Astrophysics Division, Physical Research Laboratory (PRL), Ahmedabad-380009, India}
\correspondance{jeewan@aries.res.in}
\date{1st May 2023}

\maketitle


%

\begin{abstract}
We present the timing analysis of the intermediate polar TX Col in the X-ray band using the observations made by Chandra, Swift, and Suzaku during the years 2000, 2007, and 2009, respectively. The spin, orbital, and beat periods derived from these data are consistent with the earlier findings. We found that the spin modulation was dominant during the Chandra observation, whereas both orbital and beat modulations were dominant during the Swift and Suzaku observations. These findings and past X-ray observations indicate that  TX Col is changing its accretion geometry from disc dominance to stream dominance and vice versa.  

\end{abstract}


\section{ Introduction}
Intermediate polars (IPs) are a class of magnetic cataclysmic variables where a white dwarf (WD) accretes mass from the red dwarf via Roche-lobe overflow. IPs are asynchronous systems with WD's magnetic field strength of less than 10 MG. Studying IPs is important to understand better the complex physics of accretion and magnetic fields in compact binary systems. TX Col is an IP, which is located at a distance of $923\pm26$ pc \citep{2018AJ....156...58B} with spin (\ps) to orbital (\po) period ratio of $\sim0.09$. Based on the soft X-ray detection and optical pulsation, \cite{1986ApJ...311..275T} identified TX Col as an IP. It was the first IP to show the X-ray beat period along with the spin pulse \citep{1989ApJ...344..376B} suggesting both stream-fed and disc-fed accretions. \cite{1989ApJ...344..376B} found the \ps and \po of 1911 s and 5.7 hrs using the X-ray and optical observations during the year 1984-1985. Optical photometry by \cite{1992ASPC...29..387B} in the year 1989 showed a  periodicity at half of the beat period (\psmo) of 2106 s.  Photometry of TX Col by \cite{1995ASPC...85..512S} during the year 1994 did not show the \psmo ~and its harmonic but quasi-periodic oscillations (QPOs)  at the period of 5000 s (or larger) were identified. Using ROSAT and ASCA observations, \cite{1997MNRAS.289..362N} showed that TX Col was accreting predominantly via a disc in October 1994, and a year later, it was substantially accreting via stream, making it again a disc overflow system.   In the extensive photometry spanning over $\sim$ 12 years from 1989 to 2002,  \cite{2007MNRAS.380..133M} detected $\sim 5900$ s quasi-periodic oscillations (QPOs) in the years 1990, 1994, and 2002, which they interpreted as the beating of the Keplerian period of the orbiting blobs with the spin period. In the white light photometry during the year 2002-2003, \cite{2005ASSL..332..251R} reported evidence for $\sim$ 2 h QPOs along with the large superhumps at 7.1 h  and 5.2 h, in addition to the spin, beat, and orbital periods. Recent observations from TESS show TX Col as the variable disc-overflow accretor with the presence of QPO in the period range of 3500 - 8500 s \citep{2021ApJ...912...78R,2021AJ....162...49L}. In this paper, we used the X-ray observations from Chandra, Swift, and  Suzaku to understand its accretion flow during these observations. 

\section{Observations}
X-ray observations of TX Col were taken from  Chandra, Swift, and Suzaku satellites. 
Chandra observed TX Col on July 26, 2000, at 22:37:52 UT for 50.4 ks using  Advanced CCD Imaging Spectrometer (ACIS) - I instrument \citep[][]{2002PASP..114....1W}. TX Col was observed by Neil Gehrels Swift observatory (hereafter Swift) on three occasions in December 2007 using X-ray telescope \citep[XRT;][]{2005SSRv..120..165B} and Burst Alert Telescope \citep[BAT;][]{2005SSRv..120..143B} for 5.5 to 10 ks and two occasions on October 2015 for  $<2$ ks in photon counting mode. The {\textit Suzaku} observations of TX Col were carried out on 2009, May 12 at 16:19:17 (UT) with the offset of 3.07 arcmin, using X-ray Imaging Spectrometers \citep[XIS;][]{Koyama07} and Hard X-ray Detector \citep[HXD;][]{Takahashi07}. 

We have used standard data reduction methods with updated software and calibration files for all the observations. We used the data reduction software CIAO for Chandra and HEASOFT for Swift and Suzaku. The event files from different observations were corrected for the barycentric time using the appropriate task in each satellite's data reduction procedure. Light curves of the source region were extracted with a circular region of the radius of 10", 30", and 165" for the Chandra, Swift and Suzaku observations, respectively.   A similar size of background regions around the source region was chosen for the background estimation for each observation. Finally, the source light curves were corrected for background contribution.

\section{X-ray light curves and Timing analysis}

\begin{figure}
\centering
	\subfigure[X-ray light curve]{\includegraphics[width=0.45\textwidth]{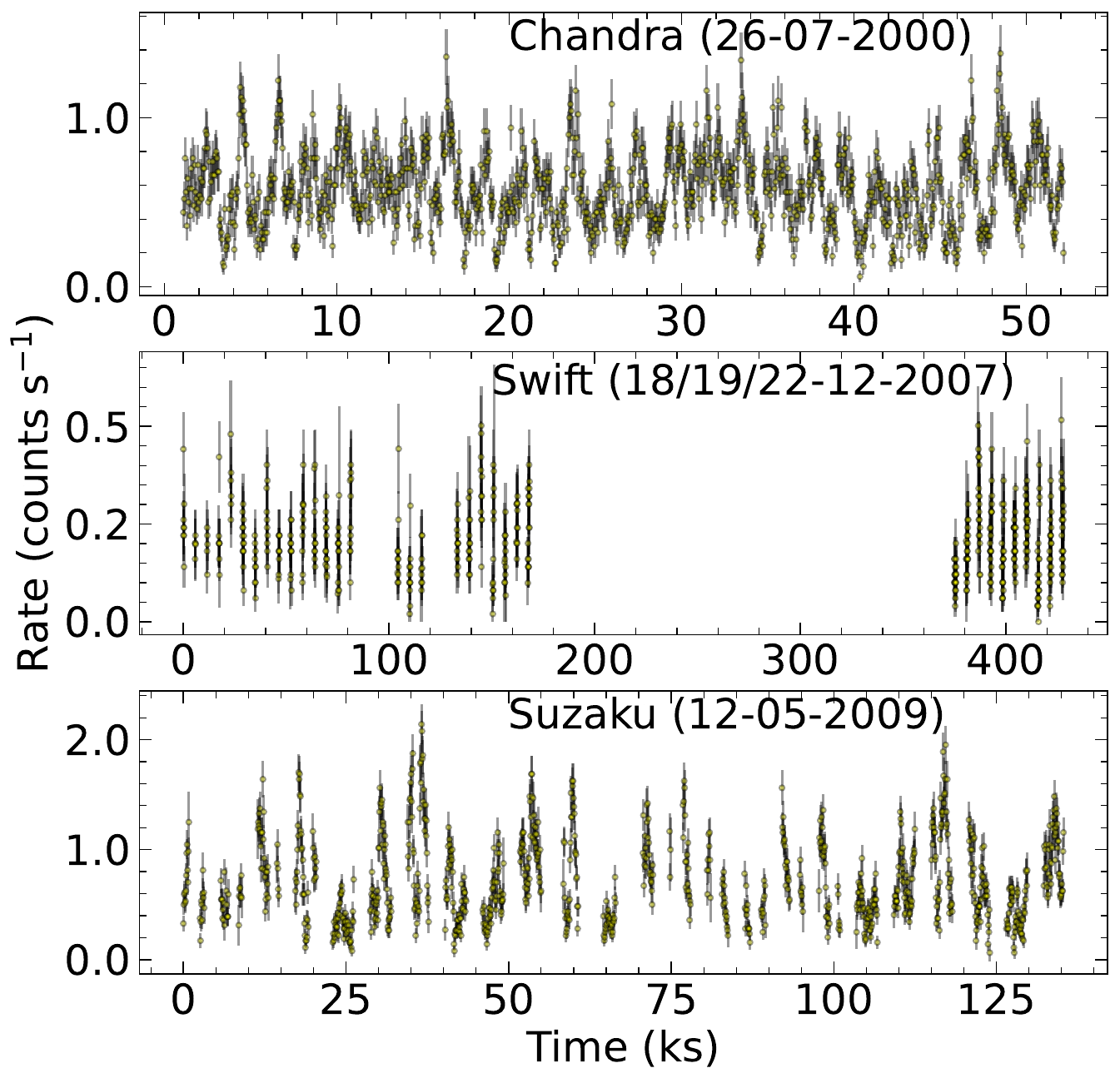} \label{fig:lc}}
	\subfigure[Power Spectra]{\includegraphics[width=0.45\textwidth]{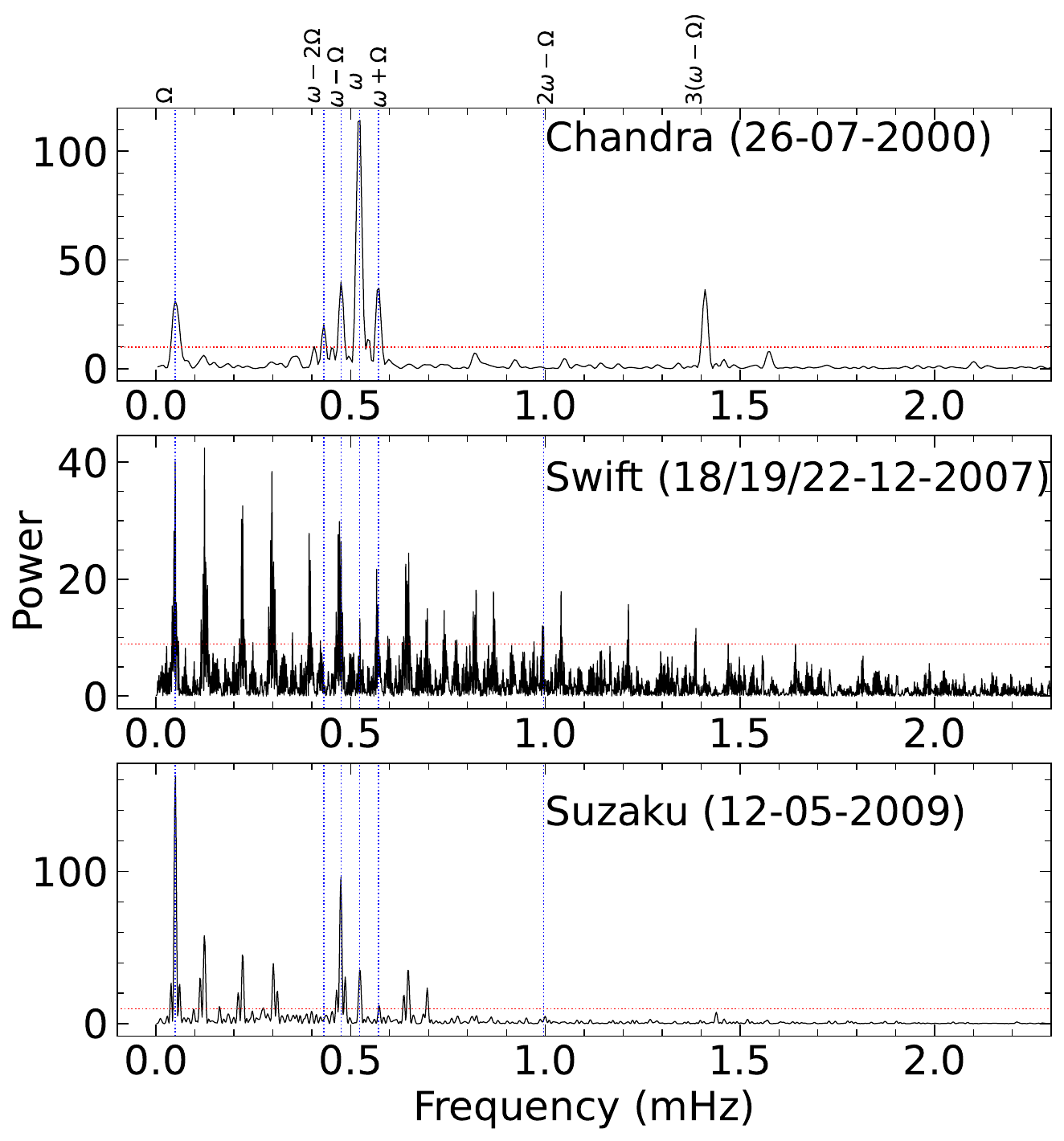}\label{fig:ps}}
\caption{X-ray light curves and corresponding power spectra of TX Col at three different epochs of observations. \label{fig:pslc}}
\end{figure}

Figure \ref{fig:lc} shows the background-subtracted X-ray light curves of TX Col as observed from ACIS-I/Chandra, XRT/Swift, and XIS01/Suzaku satellites in three different epochs. All the light curves are extracted in the energy band of 0.3-10.0 keV. The temporal binning of the Chandra, Swift, and Suzaku X-ray light curves are the 50s, 50s, and 48s, respectively.     Multi-periodic variations appear to be present in the X-ray light curves. Therefore, we have performed the timing analysis by using the Lomb-Scargle method \citep{1982ApJ...263..835S}.

Figure \ref{fig:ps} is the power spectra of X-ray light curves obtained from Chandra, Swift, and Suzaku observations. From the power spectral analysis of Chandra light curves, we identified both  \s and \ob frequencies along with several sideband frequencies above the confidence level of 99\%. Table \ref{tab:acc} shows observed frequency details. We derived the average spin period of 1909  s, orbital period of 2106 s and beat period of 5.67 hr from their different observations. These periods are very similar to those derived by the most extended high cadence optical data from TESS\citep[see][]{2021ApJ...912...78R}. In the case of Chandra's observations, we have identified sideband frequencies of \s - 2\ob, \smo, \spo, and 3(\s - \ob) above the confidence level of 99\%. Whereas for the Swift and Suzaku observations, we have found the \smo and \spo. Additionally, the 2\s-\ob frequency was also seen during the Swift observation. In the case of Chandra observations, the \s frequency was dominant, whereas the \ob frequency was dominant during the Swift and Suzaku observations.

\section{Discussion and Conclusions}
X-ray power spectra are an effective way to distinguish the accretion geometry in IPs. The commonly observed frequencies in the power spectra of IPs are  \s, \ob,  \smo, \spo, \s - 2 \ob and \s+2\ob \citep[][]{1995CAS....28.....W}. We explain the accretion phenomenon in TX Col based on \cite{1992MNRAS.255...83W}  model.   During the epoch 2000, the \s peak is dominant over \smo or any other peaks in the power spectra, opposite to that seen in the disc-less model of \cite{1992MNRAS.255...83W}. Further, \smo frequency appears to be modulated at the \ob frequency, resulting in the presence of \s-2\ob frequency in the power spectrum. This modulation is also responsible for enhancing the strength of \s peak. Further, we need to get information about the pole cap asymmetry in this system. Thus, TX Col appeared to be predominantly accreting via disc during this epoch. However, a small fraction of accretion also occurred through the stream. During the epochs 2007 and 2009, the \ob and \smo peaks are found to be more dominant than the \s peak in the power spectra. The \s peak was the weakest in the power spectra. The dominance of \ob and \smo peaks during the epoch 2007 and 2009 indicates stream-fed accretion. If  TX Col's inclination angle is low, then pure stream-fed accretion can not be considered during these epochs of observations. Also, the presence of weak \s peak suggests that a part of accretion also occurred through the disc. Thus, the system TX Col appears to accrete via disc as well as stream, being disc dominant accreator during the year 2000 and stream dominant accreator during the years 2007 and 2009.

Table \ref{tab:acc} summarises the presence of frequencies in the X-ray power spectra from different observations in the past and current analysis. We have also indicated the dominant accretion mechanism based on the timing analysis. We noticed that the power of dominant frequency changes from one observation to another. A similar result was found by using the long-term high cadence data from the TESS \citep[see][]{2021ApJ...912...78R,2021AJ....162...49L}. These X-ray observations confirm that TX Col is a variable disc-overflow system, changing its accretion mode from disc dominance to stream dominance and vice versa.   
In TX Col, the accretion geometry can change from a disc-fed to a stream-fed state and vice versa, leading to variations in the X-ray emission from the system. These changes can occur on timescales of hours to days and are thought to be due to instabilities in the disc and magnetic fields around the white dwarf. FO Aqr \citep{2020ApJ...896..116L}, V2400 Oph \citep{2019AJ....158...11J}, V902 Mon,  Swift J0746.3-1608, and UU Col \citep{2023arXiv230712962R,2022MNRAS.512.6054R} are a few other IPs in which variable disc-overflow accretion is observed. 

\begin{table}
\caption{List of frequency present in X-ray power spectra of TX Col in past and present observations.} \label{tab:acc}
\begin{tabular}{lllc}
\hline
Year & Frequencies & Dominant Frequency & Mode of accretion \\
\hline
1985$^{\star}$ & \s, \smo                              & \smo   &DOSF\\
1994$^{\dagger}$ & \ob, \s, \smo                         & \s     &DODF\\
1995$^{\dagger}$ &\ob,\s,\smo,2\s,2(\smo)                & \smo   &DOSF\\ 
2000 &\ob, \s, \smo, \spo, \s-2\ob, 3(\smo)  & \s     &DODF\\ 
2007 &\ob, \s,\smo,\spo, 2\s-\ob             & \smo and \ob &DOSF\\ 
2009 &\ob, \s, \smo, \spo                    &\smo and \ob  &DOSF\\
\hline
\end{tabular}
{\small DOSF -- Disc overflow stream fed accretion, DODF -- Disc-overflow disc fed accretion, 
$^{\star}$ \citet{1989ApJ...344..376B},
$^{\dagger}$ \citet{1997MNRAS.289..362N}}
\end{table}

\begin{acknowledgments}
This work is based on data taken from the Chandra, Swift, and Suzaku satellites.  We acknowledge the referee for careful reading of our paper.
\end{acknowledgments}

\begin{furtherinformation}

\begin{orcids}
\orcid{0000-0002-4331-1867}{Jeewan Chandra }{Pandey}
\orcid{0000-0002-4633-6832}{Nikita }{Rawat}

\end{orcids}

\begin{authorcontributions}
All authors contributed significantly to the work presented in this paper.

\end{authorcontributions}

\begin{conflictsofinterest}
The authors declare no conflict of interest.
\end{conflictsofinterest}

\end{furtherinformation}

\bibliographystyle{bullsrsl-en}

\bibliography{ref}

\end{document}